\documentclass[aps,prl,superscriptaddress,twocolumn,tightenlines,epsfig]{revtex4}

\usepackage {graphicx}

\usepackage {graphics}

\usepackage {amsmath}

\usepackage {amsfonts}

\usepackage {amssymb}

\usepackage {mathrsfs}

\usepackage{color}

\usepackage{epstopdf}

\newcommand {\be}{\begin{eqnarray}}

\newcommand {\ee}{\end{eqnarray}}

\begin{document}

\title {Metallic stripes and the universality of the anomalous half-breathing phonon  in high-T$_c$ cuprates}

\author {S.I.\ Mukhin}

\email {sergeimoscow@online.ru}

\affiliation {Theoretical Physics Department, Moscow Institute for Steel \& Alloys, Moscow, Russia}

\author {A. \ Mesaros}

\affiliation{Lorentz Institute for Theoretical Physics, Leiden University, NL}

\author {Jan \ Zaanen}

\affiliation{Lorentz Institute for Theoretical Physics, Leiden University, NL}

\author {F.V.\ Kusmartsev}

\affiliation{Department of Physics, Loughborough University, UK}


\date {\today}

\begin{abstract}

We demonstrate that the strong anomalies in the high frequency LO-phonon spectrum in cuprate
superconductors can in principle be explained by the enhanced electronic polarizability associated 
with the self-organized one dimensionality of metallic stripes. Contrary to the current
interpretation in terms of transversal stripe fluctuations, the anomaly should occur at momenta parallel to the stripes.
The doping dependence of the anomaly is naturally explained, and we predict that the 
phonon line-width and the spread of the anomaly in the transverse momentum decrease with increasing temperature while high resolution measurements should reveal a characteristic substructure to the anomaly.

\end{abstract}

\maketitle

The phonon spectrum  of the high-T$_c$ superconducting cuprates is characterized by a peculiar anomaly: halfway the Brillioun zone the 'half-breathing' $Cu$-$O$  vibration mode seems to suddenly dip down to a much lower frequency \cite{Qee,pints1,pintschovius}. The line-width reaches its maximum at a wave-vector
somewhat shifted from the frequency-dip position, while it  {\em narrows} at higher temperatures \cite{reznik}. In addition, the anomaly has a narrow intrinsic peak width as function of momentum transversal to the mode propagation direction \cite{reznik}. This is hard to explain in a conventional fermiology framework, and interpretations invoking a coupling between the phonon and purely electronic  collective modes of the stripes acquired credibility\cite{z1} by the recent demonstration that the anomaly is particularly pronounced in $La_{2-1/8}Ba_{1/8} CuO_4$ \cite{pintschovius,reznik}, a system with a well developed static stripe phase \cite{abbamonte}. Initially it was believed that the relevant stripes modes should correspond with the transversal 'meandering' fluctuations\cite{z2}, a possibility further quantified by Machida {\em et al.}\cite{machida} by computing the Gaussian fluctuations around the Hartree-Fock stripe ground state. This interpretation has however run into a serious objection: the phonon anomaly is also observed in the YBCO superconductor, where the orientation of the stripes relative to the orthorombic lattice  can be directly deduced from the anisotropy of the spin-fluctuations in the untwinned crystal. It turns out that the phonon anomaly occurs for phonon-wavevectors parallel to the stripes, at a right angle as compared to the expectations for transversal stripe modes \cite{mook}.

\begin{figure}
\begin{center} 
\includegraphics[width=0.48\textwidth]{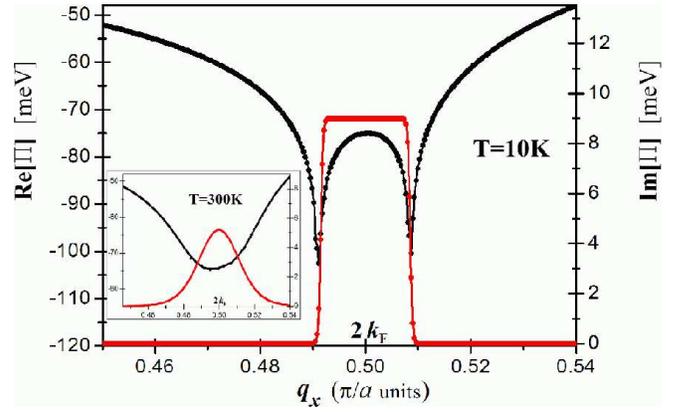}
\end{center}
\caption{Real- (black line) and imaginary (red line) part of the on-stripe electronic phonon self-energy  $\Pi(q_x,\omega)$ at $T=10$K  at a fixed (phonon)  frequency $\omega = 68$meV as function of the momentum component $q_x$ parallel to the stripes, taking for the dimensionless electron-phonon coupling a value representative for
the cuprates: $\xi=0.5$.  In the inset the results are shown at a higher $T=300$K.}
\label{1}
\end{figure}

\begin{figure}
\begin{center} 
\includegraphics[width=0.48\textwidth]{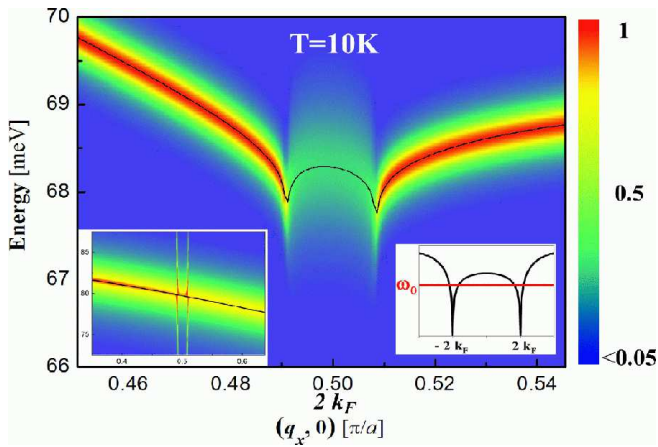}
\end{center}
\caption{False color plot of the LO phonon spectral function {\em vs} momentum in the stripe direction and energy at low temperature (10K) for 1D $\epsilon_F \sim 1$eV and the same parameters as in Fig. \ref{1}. Phonons couple merely to the 1D Fermi-gas like charge 
excitations of the metallic electron system confined in the stripes. Strong phonon damping in a momentum region 'inside' the anomaly is caused by the decay in the continuum of quasi 1D electron-hole excitations. Left inset: different behavior of the 'standard' mode coupling of the phonon with a propagating stripe collective mode (e.g., ref.\cite{machida}). Right inset:  the renormalized phonon dispersions determined by the crossing of the phonon frequency $\omega_0$
and the real part of the 1D polarization propagator of Fig. \ref{1}, see text.}  
\label{2}
\end{figure}

\begin{figure}
\begin{center} 
\includegraphics[width=8.0cm]{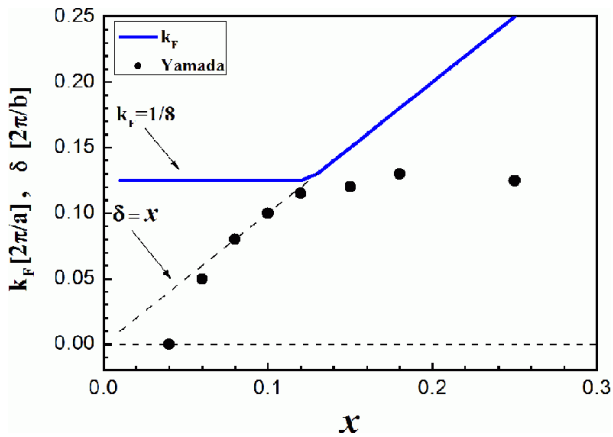}
\end{center}
\caption{The qualitatively different doping dependences expected for the characteristic wave vector of the anomaly when transversal stripe model (1) or the intra-stripe quasi 1D electron excitations (2) are responsible. In case (1) the anomaly  should follow the stripe-ordering incommensurate wave vector $\delta$ dependence on doping $x$: the famous 'Yamada plot'\cite{yamada} (black dots); (blue line) is for the softened phonon wave-vector $q=2k_F$ in case (2).}
\label{3}
\end{figure}

\begin{figure}
\begin{center}
\includegraphics[width=0.48\textwidth]{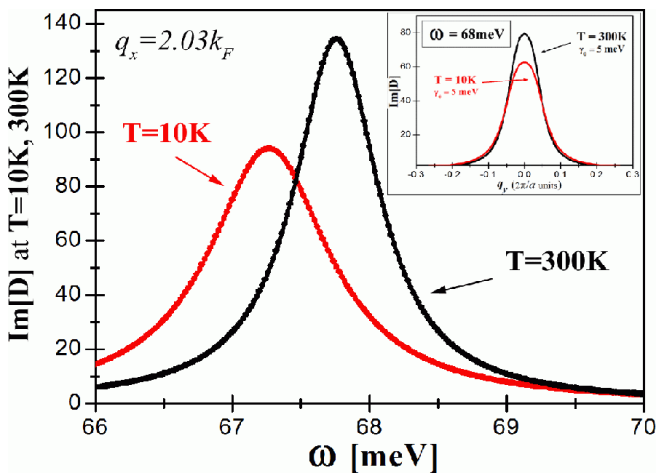}
\end{center}
\caption{Calculated phonon spectral function $ImD(q_x,q_y=0; \omega)$ dependence on frequency $\omega$, using the same parameters as in Fig.\ref{2}, at different temperatures. A temperature narrowing is predicted, accompanied by a phonon hardening at the 'dips' seen in Fig.\ref{2}. Inset: calculated $ImD(q_x,q_y; \omega)$ momentum dependence in the 'transversal' $q_y$ direction perpendicular to the stripes, at
fixed $q_x = 2k_F$ and $\omega =68$ meV. Counter intuitively, the anomaly even localizes further when temperature is raised, in accord with experiment \cite{reznik}.}
\label{4}
\end{figure}

A well known problem with the Hartree-Fock stripes is that they are 'insulating like' \cite{z3}, while cuprate stripes are quite metallic. This supports the idea that they form an electronic liquid crystal of the smectic kind as introduced by Kivelson and coworkers \cite{Kivelson}. Here the transversal modes are frozen out by commensuration effects and instead the low energy physics is governed by on-stripe compressional fluctuations. As one of us already noticed some time ago\cite{sm}, the electronic polarizability associated with the on-stripe Luttinger-liquid like physics should be strongly enhanced at the 
on-stripe $2k_F$ wave-vectors and this feature might well govern the phonon anomaly (see Fig. \ref{1}).  Embedding this 1D electron dynamics in a 2D optical phonon background is still a non-trivial exercise and here we present a first attempt to compute these matters quantitatively. We employ two simplifying assumptions: (i) we use the free fermion charge susceptibility (Lindhardt function) instead of the fully interacting 'Luttinger liquid' form, since the phonon anomaly appears to be in first instance sensitive only to the gross features of the 1D electron dynamics. (ii) More critical, we assume that at the energy of the anomaly the inter-stripe interaction effects have already diminished and this might well be an oversimplification. The outcomes are as follows. The stripe alignment problem is solved by construction, but we also arrive at some less obvious predictions: (a) the anomaly is now caused by the phonon crossing the ubiquitous continuum of 1D charge excitations centered at the
intrastripe $2k_F$ (Fig. \ref{1}). At the phonon frequency $\omega$ this continuum has a momentum width  $\Delta q \approx k_F\omega/v_c$ where $v_c$ is the electronic charge velocity and this causes a 'double dip' structure in the phonon spectral function (Fig. \ref{2}) , not unlike to what is predicted  by 
Machida {\em et al.} \cite{machida} (inset Fig. \ref{2}). However, the big difference with the propagating transversal modes of Machida is that in the 'smectic scenario'  the phonon is completely (Landau) damped in the momentum region  in between the dips: high resolution neutron scattering measurements should be able to resolve this. (b) Since the characteristic momentum where the anomaly occurs is now determined by the intrastripe electron density and not by the interstripe distance the position of the anomaly should have a doping dependence that is radically different from what is expected for transversal modes (Fig. \ref{3}). (c)  We predict that the anomalous phonon linewidth $\Delta\omega$ {\em decreases} with increasing temperature (Fig.\ref{4}).(d) Finally, although we assume that the stripe
Luttinger liquids are strictly independent, the form factors of the electron phonon interaction \cite{horsch} localize the phonon  
anomaly in 2D momentum space also in the direction  perpendicular to the stripes (Fig. {\ref 4}). Counterintuitively, we find that it contracts
further when temperature is raised (inset Fig. {\ref 4}), a behavior 
that is also observed in experiment\cite{reznik}, while it is difficult to rationalize invoking transversal stripe modes. 
  
We embed the array of parallel metallic stripes in the 2D phonon universe by considering a simplified propagator: it describes non-interacting electrons in Bloch states on a periodic array of parallel lines in two dimensions, moving freely along these lines, but with vanishing inter-line overlap of the `Wannier' functions: 

\be
G(\vec{r},\vec{r'};i\omega)=\frac{1}{N_k}\sum_{\vec{k}}\sum_{l,l'=0}^{N-1}\frac{e^{i\vec{k}(\vec{r}-\vec{r'})+iQ(ly-l'y')}}
{i\omega-\epsilon(\vec{k})}
\label{green}
\ee

\noindent
where $\vec{r},\vec{r'}$ span the the 1D $Cu$-centered stripes(lines) in real space; $\omega$  is the Matsubara frequency.  
We consider an orthorombic $Cu$-$O$ plane, with $a$($b$) being the unit cell spacing along the $x$($y$)-axis. 
The stripe metallic direction is along the $x$-axis and the stripe Umklapp
momentum $Q=2\pi/bN$ is along the $y$-axis, where $N$ ($=4$ at higher doping) is the number of unit cells in one inter-stripe (charge-density) period. The momentum $\vec{k}$ spans $N_k$ sites in the reduced orthorombic Brillouin zone $0<k_y<2\pi/bN$; $0<k_x<2\pi/a$. The electron dispersion $\epsilon(\vec{k})\approx\epsilon(k_x)$, ignoring  inter-stripe tunneling (see below). 
  
A  hole on site $\vec{r}$ inside the stripe communicates with neighboring oxygens bond-stretching displacements $u^{\vec{r}}_{\pm{i}}\equiv u_i(\vec{r}\pm\vec{i}/2)$, $\vec{i}=\vec{a},\vec{b}$ and we 
take the effective coupling between those and the Zhang-Rice singlets as introduced by  Horsch an Khaliullin \cite{horsch},

\be
H_{e-ph}=g_0\sum_{\vec{r}}\left(u^{\vec{r}}_{x}-u^{\vec{r}}_{-x}+u^{\vec{r}}_{y}-u^{\vec{r}}_{-y}\right)c^{\dag}_{\vec{r}}c_{\vec{r}}
\label{eph}
\ee

\noindent
with $g_0\approx 2eV/\dot{A}$, using the standard estimates for the charge transfer energy and hoppings\cite{sawat}.
This is actually our main step: from this Hamiltonian and  2D 'striped' electron propagator Eq. (\ref{green}) it is straightforward to calculate the self-energy part of the dynamical matrix associated with
the $Cu$-$O$ plane:

\be
\Lambda^{E\alpha,\beta}_{x,y}(\vec{q},\omega)=\omega_0\Pi(q_x, \omega)\begin{pmatrix}
s^2_{q_x} & s_{q_x}s_{q_y} \\
s_{q_x}s_{q_y} & s^2_{q_y}
\end{pmatrix}
\label{det}
\ee
\noindent
where $\omega_0$ is bare phonon frequency, $s_{q_x}=\sin{q_xa}/{2}$, $s_{q_y}=\sin{q_ya}/{2}$, and $\alpha,\beta=1,2,3$ enumerate ions in the in-plane $Cu$-$O$ unit cell: 
this particular form of the electron-phonon form-factors follows immediately from the tight binding Hamiltonian Eq. (\ref{eph}), implying actually a substantial dependence on the momentum $q_y$ perpendicular to the stripes. ${\Pi}(q_x,\omega)$ corresponds with the polarization propagator of the (in principle, interacting) on-stripe Luttinger liquid,  depending  on the momentum component $q_x$ along the stripes, while its fermion
lines are given by the propagator Eq.(\ref{green}). 

This self-energy has to be added to  the bare (undoped) $ab$-plane ionic $6\times 6$ dynamic matrix $\Lambda^{I\alpha,\beta}_{i,j}(\vec{q},\omega)$ 
of the  orthorombic planes, constructed to be in close agreement with the experimental data \cite{pintschovius} for the in-plane bond-stretching LO phonon modes in the undoped cuprates. The phonon spectra $\omega_{\sigma}(\vec{q})$ associated with polarizations ${\vec{e}}^\alpha_{\vec{q},\sigma}$ are obtained by the diagonalization of the total dynamic matrix \cite{falter}.
From these solutions the phonon spectral functions given by the imaginary part of the phonon propagator $D(\vec{q}, \omega)$ are
obtained. Notice that the $\Lambda^E$'s are in general complex quantities,
with the effect that  $\omega_{\sigma}(\vec{q})$ acquires an imaginary part representing the phonon damping.

\noindent
We use for ${\Pi}(q_x,\omega)$ the well known 1D Lindhardt function\cite{weger},
\be
Re{\Pi}(q,\omega)=-\dfrac{\omega_0\xi}{4\pi q\tau}\int_{0}^{\infty}{ln\left|\dfrac{\Delta_{+}}{\Delta_{-}}\right|}
{ch^{-2}\left(\dfrac{p^2-1}{2\tau}\right)}pdp \; &&\label{1df}\\
Im{\Pi}(q,\omega)=\dfrac{\omega_0\xi}{8}\left[th\dfrac{\omega+2(q-2)}{4\tau}+th\dfrac{\omega-2(q-2)}{4\tau}\right]\;&&
\label{im1df}
\ee
\noindent 
here $\Delta_{\pm}=(2p\pm q)^2q^2-\omega^2$, all momenta and energies are measured in units of the Fermi-gas parameters $k_F$ and $\epsilon_F\equiv k_{F}^2/2m\sim 1$eV; $\tau=k_BT/\epsilon_F$ is a dimensionless temperature. The dimensionless electron-phonon coupling constant $\xi=g^2_0/K\epsilon_F\sim 0.5$ is representative for cuprates with $K\approx 25eV/\dot{A}^2$ the lattice force constant \cite{horsch}. 

The effect of this 1D polarizability on the phonons follows from  the behavior of $\Pi (q_x, \omega)$ at the phonon-frequency $\omega= \omega_0$ as function of $q_x$ (Fig. 1). The continuum of charge excitations in a 1D fermi-gas is the well known fan in the momentum-frequency plane, centered at $2k_F$ at $\omega =0$ and bounded by $2k_F \pm \omega / v_F$ at finite frequency. For non-interacting electrons the spectral function ($Im \Pi$) is just the box of Fig. (\ref{1}), while in the presence of interactions the spectral weight will pile up at the edges. Since $\Pi$ is proportional to the phonon self-energy, the phonon spectral function, Fig. (\ref{2}), indicates that the phonon dispersion is pushed downwards when it approaches the edges of the quasi 1D electron-hole continuum from either side, to broaden strongly when it enters the continuum. This is markedly different from the result based on a mode coupling between the phonon and a propagating mode as for instance discussed by Machida {\em et al.} \cite{machida} as shown in the inset of 
Fig. (\ref{2}): in this case there is no phonon damping and the intensity is just distributed over the propagating modes subjected to an avoided level crossing.  

The results shown in Fig. (\ref{2}) suggest that it is unproblematic to give a quantitative account of the observed electron-phonon effects in terms of the metallic stripe electrons, provided that flatness of their band dispersion $\epsilon_{\perp}(k_y)$ perpendicular to stripes is bounded from above by $\sim\omega_0$. The numerical LDA$+U$ calculations \cite{anisimov} support this assumption, suggesting $\epsilon_{\perp}\sim 15meV\approx 0.2\omega_0$. 
Alltogether, this  success is not all that remarkable since the 'propagating mode' and the 1D metallic stripe scenario's give in this regard the same answer: by concentrating the charge excitations in a small kinematic region (as compared to conventional fermiology) one is dealing in essence with an avoided mode crossing story and because this is about resonance small numbers can give big effects.
 
A distiction between the effects of the transversal stripe modes and the internal 1D-like fermionic excitations on the phonon anomaly should be revealed by the different doping dependences of the locus of the phonon anomaly in momentum space. The transversal fluctuations emerge at the stripe ordering wave vectors and these should follow the famous Yamada plot\cite{yamada}, correlating the stripe ordering wavevectors $\delta$ with doping $x$ (Fig. \ref{3}), such that at low dopings the anomaly should live at a wavevector $q_{a} \sim 1/x$. On the other hand, dealing with the quasi 1D modes the locus of the anomaly is determined by the on-stripe electron density and according to the Yamada plot this stays constant  ('half-filled')  at $q_{a} = 2k_F = \pi / 2a$ up to $x_c= 1/8$, while at higher dopings it should follow
$q_a = 2k_F\propto x-x_c$ because the on stripe hole density is increasing (the blue line in Fig. \ref{3}). 
 Experimentally the locus of the anomaly in $k$-space is conspicuously doping independent\cite{pintschovius}, actually arguing strongly against the transversal mode. It would be interesting to find out if the 'center of mass' of the anomaly does shift at higher dopings. 

\noindent
Our theory yields a rational for the observed gross temperature dependences of the anomaly \cite{reznik}. The rather counterintuitive narrowing of the frequency width with increasing temperature follows naturally from the 1D polarization propagator decrease, particularly: $Im{\Pi}\propto \omega/T$ (Fig. \ref{1}). The effect of this change on the phonon-spectral function is shown in Fig. (\ref{4}): a substantial narrowing occurs at higher temperature. Moreover, right at the 'dips' a substantial phonon hardening occurs since the phonon positions at these momenta are most sensitive to the details of the real part of the self energy. 

\noindent
The phonon anomaly behavior in the 'transversal'
$q_y$-direction (inset, Fig. \ref{4}) follows from our analysis of the expression for the phonon spectral function $ImD(q_x,q_y; \omega)$ showing a substantial $q_y$ dependence  close to the anomaly  due to the form-factors in Eq. (\ref{det}):

\be
ImD(q_x,q_y;\omega)\approx\dfrac{\omega_0\Delta_{q_x,q_y}Im{\Pi}_{q_x}}{4(\tilde{\omega}_{q_x,q_y}-\omega)^2+\Delta_{q_x,q_y}^2Im^2{\Pi}_{q_x}}
\label{qy}
\ee
\noindent 
where $\Delta_{q_x,q_y}\equiv s^2_{q_x}+s^2_{q_y}$, and $\tilde{\omega}_{q_x,q_y}\approx \omega_0+0.5\Delta_{q_x,q_y}Re{\Pi}_{q_x}$ is the renormalized optical phonon mode frequency. The width of the Lorentzian with respect to $q_y$ at $\omega=\tilde{\omega}_{q_x=2k_F,q_y=0}$ is:
$\delta q_y\approx 2/a\sqrt{|Im\Pi/Re\Pi|}\sim 0.1\times 2\pi/a$, assuming a flat bare mode dispersion.  Given the ratio under the square root, the width $\delta q_y$ decreases when the temperature is raised. 
 
In summary, we have analyzed a minimal- and likely too oversimplifed model dealing with the `Luttinger liquid'-like excitations coming from the electrons confined in stripes interacting with optical lattice phonons. The gross features of the phonon anomaly as measured experimentally are consistent with the workings of a quasi 1D array of metallic intrastripe Luttinger liquids. More importantly, although it will take a significant investment, it does make sense to measure the phonon anomaly with a much higher resolution than has accomplished up to now. Such measurements should reveal the internal structure of the anomaly and as we showed in this paper, from this information much can be learned regarding the still quite mysterious organization of the electron systems in high T$_c$ superconductors. 

\noindent 
We acknowledge financial support by 
the Netherlands foundation for fundamental research of Matter (FOM)
and the dutch organization for scientific research (NWO). S.I.M. acknowledges support in the form of  a Royal Society (UK) grant for international incoming short visits at Loughborough University.


\begin{thebibliography}{}

\bibitem{Qee}R.J.McQueeney,Y. Petrov, T. Egami, M. Yethiraj, G. Shirane, and
Y. Endoh, Phys. Rev. Lett.{\bf{82}}, 628 (1999).
\bibitem{pints1}L. Pintschovius, W. Reichardt, M. Klaser, T. Wolf, and H.v. Lohneysen, Phys. Rev. Lett. {\bf{89}}, 037001 (2002).
\bibitem{pintschovius}L. Pintschovius, Phys. Stat. Sol. (b){\bf{242}}, 30 (2005); D. Reznik {\it{et al.}}, Nature, {\bf 440}, 1170 (2006).
\bibitem{reznik} D.Reznik, L. Pintschovius, M. Fujita {\em et al.}, JLTP No.2 (2007). 
\bibitem{z1}J.Zaanen, Nature {\bf{440}},1118(2006).
\bibitem{abbamonte}P. Abbamonte {\it{et al.}} Nature Physics {\bf{1}}, 155 (2005).
\bibitem{z2}J. Zaanen, M.L. Horbach, and W.van Saarloos, Phys. Rev. B {\bf{53}}, 8671(1996);
H.Eskes,O.Y. Osman,R. Grimberg,W. van Saarloos,J. Zaanen,Phys. Rev. B {\bf{58}}, 6963(1998).
\bibitem{machida}E. Kaneshita, M. Ichioka, and K. Machida, Phys. Rev. Lett. {\bf{88}}, 115501 (2002).
\bibitem{mook}H.A. Mook, P. Dai, F. Dogan, and R.D. Hunt, Nature (London){\bf{404}}, 729 (2000).
\bibitem{z3}J.Zaanen and O.Gunnarsson, Phys.Rev.B {\bf{40}},7391(1989); J.Zaanen and A.M.Oles, Ann.Physik {\bf{5}},224(1996).
\bibitem{Kivelson} S.A. Kivelson, E. Fradkin, V.J. Emery, Nature vol. 393, 550
  (1998); E. Fradkin and S.A. Kivelson,Phys. Rev. B{\bf{59}},8065 (1999);D.G. Barci, E. Fradkin,S.A. Kivelson, V. Oganesyan,Phys. Rev. B{\bf{65}},245319 (2002); S.A. Kivelson, I.P. Bindloss, E. Fradkin {\it{et al.}},Rev. Mod. Phys.{\bf{75}}, 1201 (2003); E. Arrigoni,E. Fradkin and S.A. Kivelson, Phys. Rev. B{\bf{69}}, 214519 (2004).
\bibitem{sm}S.I. Mukhin, ``Phonon softening: ``keyhole'' into dynamic stripe-phase'', 
cond-mat/0507294 (2005).
\bibitem{horsch} G. Khaliullin and P. Horsch, Physica C {\bf{162-164}},462 (1989); P. Horsch and G. Khaliullin,Physica B {\bf{359-361}},620 (2005).
\bibitem{sawat}F. Barriquand, G.A. Sawatzky,Phys. Rev. B {\bf{50}},16649 (1994).
\bibitem{falter} C. Falter, Physics Reports {\bf{164}}, 1 (1988). 
\bibitem{weger}B. Horovitz, H. Gutfreund and M. Weger, Phys. Rev. B, {\bf{12}},3174 (1975).
\bibitem{anisimov} V.I. Anisimov {\it{et al.}},Phys. Rev. B, {\bf{70}}, 172501 (2004).
\bibitem{yamada} K. Yamada {\it{et al.}}, Phys. Rev. B {\bf{57}},6165 (1998). 

\end{thebibliography}
\end {document}